\documentclass[a4paper,11pt]{article}
\usepackage[dvips]{graphicx}
\usepackage{pslatex,pstcol}
\usepackage{psfrag,pifont,amsmath,amssymb,epsfig}
\newcommand{\RRxi}{\frac{R}{\xi}}

\newcommand{\rxi}{\frac{r}{\xi}}

\newcommand{\dgam}{\Delta\gamma}
\newcommand{\gamlv}{\gamma_{lv}}
\newcommand{\gamsv}{\gamma_{sv}}

\newcommand{\gamsl}{\gamma_{sl}}
\newcommand{\lqll}{\bar{\ell}}

\newcommand{\deltat}{\Lambda(T)}

\newcommand{\Mp}{M^*}

\newcommand{\beg}{\begin{equation}}
\newcommand{\eeq}{\end{equation}}
\newcommand{\beqa}{\begin{eqnarray}}
\newcommand{\eeqa}{\end{eqnarray}}
\newcommand{\bal}{\begin{align}}
\newcommand{\eal}{\end{align}}
\newcommand{\bsub}{\begin{subequations}}
\newcommand{\esub}{\end{subequations}}

\renewcommand{\eqref}[1]{Eq.~(\ref{#1})}

\title{Surface and bulk melting of small metal clusters }
\author{Johan Chang$^{1,2}$ and Erik Johnson$^{1,3}$\\[2mm]
\text{\small{$^1$Niels Bohr Institute, University of Copenhagen, Universitetsparken 5, DK-2100 Copenhagen , Denmark}}\\
\text{\small{$^{2}$Department of Quantum Engineering, Nagoya University, Nagoya 464-8603, Japan}}\\
\text{\small{$^3$Department of Materials Research Ris\o\ National Laboratory, DK-4000 Roskilde, Denmark}}}
\date{}

\begin{document}
\maketitle
\begin{flushleft}
\text{\bf{ABSTRACT}}
\end{flushleft}
We present an analytical solution to the two-parabola Landau model, applied to melting of metal particles with sizes in the nanoscale range. The results provide an analytical understanding of the recently observed pseudo-crystalline phase of nanoscale Sn particles. Liquid skin formation as a precursor of melting is found to occur only for particles with radii, greater than an explicitly given critical radius. The size effect of the melting temperature and the latent heat  has been calculated and quantitative agreement with experiments on Sn particles was found. 
\section*{Introduction}
Among the thermodynamical effects that depend on the size of the system, the depression of melting temperature when one or more dimension of a crystal is reduced to nanometer scale, is one of the best known. The melting point depression has been intensively investigated since it was experimentally discovered by Takagi \cite{takagi} in 1954.
 Melting of small metal particles \cite{allen1,buffat,oshima,garrigos} as well as semiconductor particles \cite{goldstein} at temperatures below the bulk melting temperature is now experimentally well established. 
The phenomenon was already predicted theoretically by Pawlow \cite{pawlow} in 1909.  Since then several models have been proposed, for a review see \cite{borel,jesser}. Commonly these models  predict a linear relationship between the depressed melting temperature and the inverse radius of the particle owing to the increase in the role of the surface free energy.

More recently, studies of ultra-fine particles have indicated that small scale systems may exhibit quantum-size effects that modify the electronic states. Observations of morphology and internal structures of ultra-fine particles show the presence of a pseudo-crystalline phase (where the particles are continuously fluctuating between different structures) \cite{oshima,iijima}. The pseudo-crystalline phase has been discussed theoretically by \cite{ajayan}.

The quantum-size effects occur when the number of surface atoms $N_s$ are comparable to the number of bulk atoms $N$. For a spherical particle with radius $R$ we have from \cite{kofman}:
\beg
\frac{N_s}{N}\approx\frac{3a}{2R},
\eeq
where $a$ is the atomic spacing of the crystal. A typical metal particle with  radius  $R=5 nm$ will therefore have about $10\%$ of the atoms at the surface.

 Premelting of  surface layers has recently been studied intensively both theoretically and experimentally. Applying Landau theory to a semi-infinite system, \cite{lipowsky1,lipowsky} have shown that surface melting may induce critical phenomena. Later, \cite{pluis,pluis1,pluis2} have studied surface melting of semi-infinite systems both experimentally and theoretically, and good agreement was found between experiments and the two-parabola Landau model. Surface melting of finite systems has been studied experimentally by means of electron transmission microscopy (TEM) \cite{kofman,saka,johnson} and X-ray techniques \cite{xray1,xray} however only few theoretical studies have been reported \cite{richard}.

To address the question of thermodynamical size effects and surface melting of finite size systems 
 we take an analytic approach, solving the two-parabola Landau model suggested by \cite{pluis} that was analyzed numerically for finite systems by \cite{sakai}.
\section*{Model}
The Landau free energy functional $F[M(R)]$ for a spherical particle of radius $R$ described by the ordering parameter $M(r)$ is given by \cite{sakai}
\beg\label{eq:freeenergy}
F[M(R)]=4\pi R^2 f_s(M)+4\pi\int_0^R r^2\left\{f(M(r))+\frac{J}{2}\left(\frac{dM}{dr}\right)^2\right\}dr,
\eeq
where the ordering parameter $M=1$ describes the perfect crystal and $M=0$ describes pure liquid. The term $f_s(M(R))$ is the free energy contribution from the surface per unit area and the term $f(M(r))$ is the free energy per unit volume for homogeneous bulk material. The second term of the integrand is the gradient in the order parameter and $J$ is assumed to be constant \cite{sakai}. 
The free energy function $f(M(r))$ is assumed to be polynomial and given by a two-parabola expression 
\begin{equation}
f(M)=
        \begin{cases}
                \frac{\alpha}{2}M^2(r)+\deltat& \text{for $M<\Mp$}\\
                \frac{\alpha}{2}(1-M(r))^2& \text{for $M>\Mp$}
        \end{cases}
\end{equation}
where $\alpha$ is a material dependent parameter and $\Mp$ is the intersection points of the two parabola:
\beg
\Mp=\frac{1}{2}-\frac{\deltat}{\alpha}.
\eeq
$\deltat$ is the difference in homogeneous bulk free energy per unit volume between the liquid and the solid phase at temperature $T$. Approximating to first order in the temperature
\beg\label{eq:deltat}
\deltat\approx L_b\frac{T_m-T}{T_m}
\eeq
where $L_b$ is the bulk latent heat of melting per unit volume and $T_m$ is the bulk melting point,
the surface energy term can be expressed as \cite{lipowsky}
\beg
f_s=\frac{\alpha_s}{2}M^2(R)+\gamlv,
\eeq
where $\alpha_s$ is a material dependent parameter and $\gamlv$ is the liquid-vacuum interfacial energy per unit area. The parameters $\alpha_s,\alpha$ and $J$ are related to the interfacial energies $\gamsl,\gamsv,\gamlv$ and $\Delta\gamma=\gamsv-\gamsl-\gamlv$, where $\gamsl$ and $\gamsv$ are the solid-liquid and solid-vacuum interfacial energies per unit area respectively. From \cite{pluis} we have: 
\beg\label{eq:xical}
\xi=\sqrt{\frac{J}{\alpha}},
\eeq
$\xi$ can be identified with the correlation length in the liquid state. The following relations are also given in \cite{pluis}:
\beg
\alpha=\frac{4\gamsl}{\xi} \qquad J=4\gamsl\xi  \qquad \alpha_s=\frac{1+\frac{\Delta\gamma}{\gamsl}}{1-\frac{\Delta\gamma}{\gamsl}}4\gamsl.
\eeq
Further $\kappa$ is defined as:
\beg
\kappa=\frac{J}{\xi\alpha_s}
\eeq
\subsection*{Solution}
The order parameter $M(r)$ that minimizes the total free energy $F[M(R)]$ is found by applying the variational principle $\delta F[M]/\delta M=0$. The Euler-Lagrange condition then implies:
\beg
\frac{d^2M}{dr^2}+\frac{2}{r}\frac{dM}{dr}+\frac{1}{\xi^2}(1-M)=0.
\eeq

The solution is given by a sum of the first and second kind of modified spherical Bessel functions of zero order:
\beg
1-M(r)=A\frac{\exp(-r/\xi)}{r/\xi}+B \frac{\sinh(r/\xi)}{r/\xi},
\eeq
where $A$ and $B$ are constants. The singularity at $r=0$, implies $A=0$ while $B$ is determined by the surface condition \cite{lipowsky}:
\beg
J\frac{dM}{dr}(R)=\frac{\partial f_s}{\partial M(R)}.
\eeq
The pure crystal phase then has an order parameter profile given by:
\beg\label{eq:dry}
M_{CRY}(r)=1-\frac{1}{1+\kappa\left[\coth(\RRxi)-\frac{\xi}{R}\right]}\frac{R}{r}\frac{\sinh(\rxi)}{\sinh(\RRxi)}.
\eeq
\begin{figure}
\centering
\psfrag{R=0.5xi}{\text{\tiny{$R=0.5\xi$}}}
\psfrag{R=2xi}{\text{\tiny{$R=2\xi$}}}
\psfrag{R=5xi}{\text{\tiny{$R=5\xi$}}}
\psfrag{R=10xi}[b]{\text{\tiny{$R=10\xi$}}}
\psfrag{R=50xi}{\text{\tiny{$R=50\xi$}}}
\psfrag{485 K}{\text{\tiny{485 K}}}
\psfrag{Liquid}{\text{\tiny{Liquid}}}
\psfrag{r/R}[t]{\text{\small{r/R}}}
\psfrag{Order}[b]{\text{\tiny{Order Parameter $M$}}}
\psfrag{OrderM}[B]{\text{\tiny{Order Parameter $M_{CRY}$}}}
\psfrag{440 K}{\text{\tiny{440 K}}}
\psfrag{480 K}{\text{\tiny{484 K}}}
\psfrag{Solid}{\text{\tiny{Solid}}}
\psfrag{Quasi Liquid}{\text{\tiny{QLL}}}
\psfrag{b}{\text{\small{b}}}
\psfrag{a}{\text{\small{a}}}
\begin{center}
\includegraphics[width=0.9\textwidth]{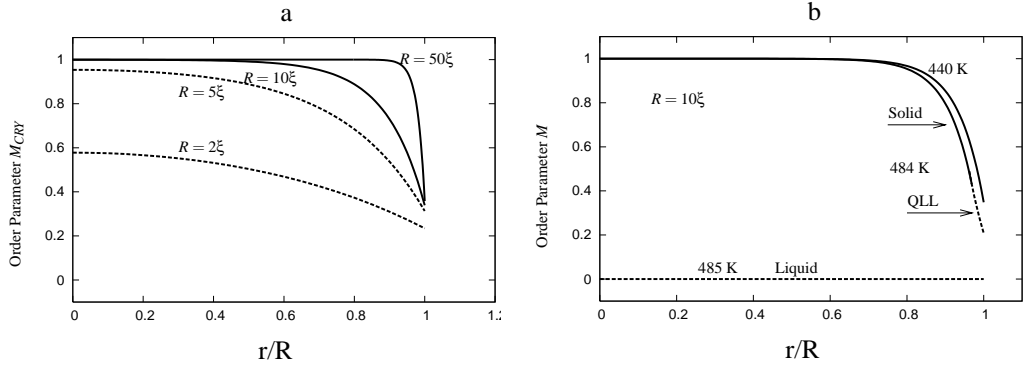}
\end{center}
\caption{{\protect\footnotesize{(a) Order parameter profile $M_{CRY}$ for Sn particles with different sizes. When $R<5\xi$ a pseudo-crystalline phase occurs. (b) The order parameter profile for a Sn particle with radius $R=10\xi$ as function of temperature.}}}
 \label{mprofile}
\end{figure}
The crystal order parameter profile $M_{CRY}$, equation \ref{eq:dry}, has been plotted for tin particles with different radii $R$ in figure \ref{mprofile}a. Naturally the atoms near the surface are less ordered than the interior of the particle. Surprisingly, significant disorder of the interior of the particle is seen for particles with radius $R<5\xi$.
 We interpret this phenomenon to corresponds to the pseudo-crystalline phase observed for ultra-fine tin particles by Oshima and Takayanagi \cite{oshima}. The pseudo-crystalline phase has been observed for tin clusters with $R<2.5nm$. Clusters where $2.5nm<R<3.5nm$ were mainly observed to be crystalline while a small fraction were still in the pseudo-crystalline phase. The correlation length $\xi$ is typically of the order of $1 nm$, thus giving good agreement between the model and the experiment of \cite{oshima}. The origin of the pseudo-crystalline phase is, of course, the increasing importance of the number of surface atoms.

For the sake of simplicity the order parameter profile $M(r)$ including a quasi liquid layer (QLL) phase $M_{QLL}$ is given in the appendix. Figure (\ref{mprofile}b) shows the temperature-dependence of the order parameter profile $M(r)$ for a Sn particle with radius $R=10\xi$. The solid curves are for $M(r)>\Mp$  and represent the solid state. The dashed curves are for $M(r)<\Mp$ and represent the liquid and quasi liquid states. It is seen that the order profile $M(r)$ does not change significantly due to the presence of the quasi liquid layer (QLL) and that the quasi liquid layer remains thin until the complete melting.
\section*{Size dependence of the latent heat of fusion}
In classical thermodynamics the latent heat of fusion is assumed to be constant and independent of the size of the system. Recent experiments \cite{lai} and MD simulations \cite{md,md1}, however indicate that the latent heat of fusion may be size-dependent. From the size effects of the order parameter profile M(r), see figure (\ref{mprofile}), it is natural to expect a decrease in the latent heat of fusion as the particle size decreases. However it is not obvious how the order parameter profile can be linked to the latent heat of fusion.\\ 
We suggest the following definition:
\beg\label{eq:defi}
L_m(R)\equiv L_b \frac{4\pi\int_0^R r^2 M_{CRY}^2(r)dr}{4\pi\int_0^R r^2dr}
\eeq
\begin{figure}
\begin{center}
\psfrag{Radius R[nm]}[t]{\text{\small{Radius R in nm}}}
\psfrag{Temperature T[K]}[b]{\text{\small{Temp. in K}}}
\psfrag{R}[t]{\text{\small{Radius R in nm}}}
\psfrag{L}[B]{\text{\small{$L_m(R)$ [J/g]}}}
\psfrag{L_b}{\text{\small{$L_b=59.2J/g$}}}
\includegraphics[width=0.56\textwidth]{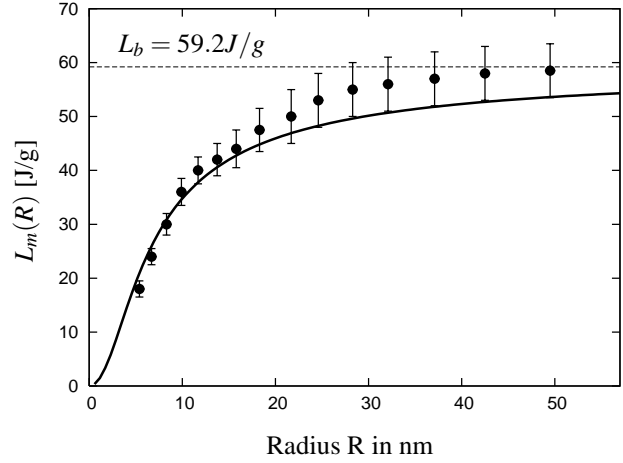}
\end{center}
\caption{\footnotesize{Size dependence of the latent heat of fusion for Sn-particles, the datum-points are taken from \cite{lai} and the solid line is calculated from equation \ref{eq:lb}, with $\xi=1.5 nm$, $\gamsv=654$mJ/m$^2$, $\gamlv=570$mJ/m$^2$ and $\gamsl=66$mJ/m$^2$.
}}\label{fig:snlb}
\end{figure} 
where $L_b$ is the bulk latent heat of fusion. For a system with perfect crystal structure ($M_{CRY}(r)=1$) the latent heat of fusion $L_m(R)$ will be equal to the bulk latent of heat of fusion $L_b$. However the crystal structure of a finite size system is not always perfect, see figure (\ref{mprofile}). Surface effects  makes the latent heat of fusion decrease with decreasing size because the order parameter $M(r)$ is less than one as the particle surface is approached. With use of the crystal order parameter profile $M_{CRY}$, equation (\ref{eq:dry}), the  evaluation of the integral in equation (\ref{eq:defi}) gives:
\beg\label{eq:lb}
L_m(R)=L_b\left\{1-\frac{3}{2\sinh^2(R/\xi)}\frac{1}{\Gamma^2}+\frac{\coth(R/\xi)[1-4\Gamma]}{\Gamma^2}\frac{3\xi}{2R}+\frac{6\xi^2}{\Gamma R^2}\right\},
\eeq
where
\beg
\Gamma=1+\kappa[\coth(R/\xi)-\xi/R].
\eeq
Note that in the thermodynamical regime, $R\gg\xi$, the latent heat of fusion $L_m(R)$ is size independent, thus $L_m(R)=L_b$.
Throughout this paper the interfacial energies for Sn are taken as  $\gamsv=654$mJ/m$^2$, $\gamlv=570$mJ/m$^2$ and $\gamsl=66$mJ/m$^2$ \cite{pluis}
while the bulk latent  heat of fusion for Sn is taken as $\tilde{L}_b=59.2$ $J/g$ \cite{hand} or $L_b=\tilde{L}_b\rho_{\ell}=411.2MJ/m^3$, where $\rho_{\ell}$ is the liquid density. The correlation length $\xi$ is estimated from \cite{pluis}:
\beg
\xi\approx\frac{5.7\gamsl}{\rho_{l}k_BT_m},
\eeq
where $\rho_{l}=N/V$ is the particle number density of the liquid state and $k_B$ is the Boltzmann constant. For Sn  $\rho_{l}=354\times 10^{26} m^{-3}$ and $T_m=505K$ thus the correlation length $\xi$ for Sn can be estimated to $\xi=1.5 nm$.
With these parameters equation (\ref{eq:lb}) is compared with a experiment done on supported free Sn particles \cite{lai}, in figure (\ref{fig:snlb}). Within the experimental uncertainty there is good agreement between the experimental and calculated data.
\subsection*{Sharp bulk melting}
It is known that ultra fine particles can not exhibit a strict phase transition. Finite-size scaling analysis shows that a first order phase transition of a finite system is broadened over an interval:
\beg
\frac{\Delta T_c}{T_c}\sim\frac{k_B T_c}{L N}
\eeq
where $k_B$ is the Boltzmann constant, N is the number of particles and $L$ is the latent heat per atom \cite{imry}. This broadened interval has also been found from quantum-statistical considerations by \cite{berry}. For a typical metal particle with a radius $R=5 nm$ we find that $\Delta T_c\sim 0.1K$, thus the transition is rather sharp. 
\begin{figure}
\begin{center}
\psfrag{R}[t]{\text{\small{Radius R in $nm$}}}
\psfrag{T}{\text{\small{$T_c$ in K}}}
\psfrag{Tm}{\text{\small{$T_m=505$K}}}
\includegraphics[width=0.56\textwidth]{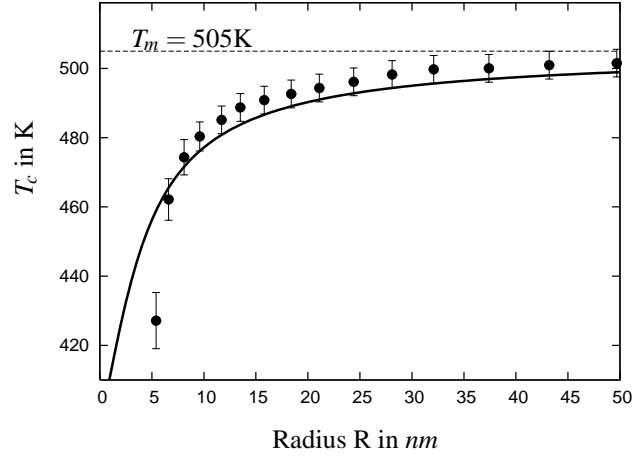}
\end{center}
\caption{\footnotesize{Size dependence of the melting temperature of supported free  Sn-particles, the datum-points are taken from \cite{lai} and the solid line is calculated from equation \ref{eq:nosurf}, with $\xi=1.5 nm$, $\gamsv=654$mJ/m$^2$, $\gamlv=570$mJ/m$^2$ and $\gamsl=66$mJ/m$^2$.}}\label{fig:snmelt}
\end{figure} 
In figure \ref{mprofile}b, it is  seen that the order parameter profile $M(r)$for a particle with dry surface does not change significant by surface melting. Therefore, it is a good approximation to consider melting of a small particle as a first order transition from a dry particle to a  liquid particle. The melting transition occurs when the Landau free energy of the solid and the liquid phase equals.
 The critical melting temperature $T_c$ is therefore found by solving $F[M_{CRY}]=F[M=0]$ where $F$ is given in equation (\ref{eq:freeenergy}):
\beg\label{eq:nosurf}
\frac{T_m-T_c}{T_m}=\frac{6\gamsl}{R L_b}\frac{\coth(R/\xi)-\xi/R}{1+\kappa\left[\coth(R/\xi)-\xi/R\right]}.
\eeq
In figure \ref{fig:snmelt} the melting temperature $T_c$ of supported free Sn particles is compared with the model. With the same interfacial energy values as mentioned before good agreement was found except for the two smallest particle sizes. \\
Remark that in the thermodynamical regime, $R\gg\xi$, the  melting temperature $T_c$ simplifies to the classical thermodynamical result: 
\beg\label{eq:firsto}
\frac{T_m-T_c}{T_m}=\frac{3(\gamsv-\gamlv)}{L_b}\frac{1}{R}
\eeq
where the depressed melting temperature $(T_m-T_c)/T_m$ is linear proportional with the inverse radius $R$. In the opposite regime, $R\ll\xi$, it was found from equation \ref{eq:nosurf} that the melting temperature $T_c$ is size independent:
\beg
\frac{T_m-T_c}{T_m}=\frac{2\gamsl}{L_b \xi}.
\eeq 
In this regime, a quantum mechanical model is necessary to describe the phenomenon properly. However the Landau model should be valid in the intermediated regime $R\approx\xi$.
\section*{Surface and bulk melting}
\begin{figure}
\begin{center}
\psfrag{gam}{$\dgam$}
\psfrag{R}{R}
\psfrag{Rp}{$R_c$}
\psfrag{Rm}{$R_c^-$}
\psfrag{surface no-melting}{\text{\small{Surface non-melting}}}
\psfrag{Surface no-melting}{\text{\small{Surface non-melting}}}
\includegraphics[width=0.56\textwidth]{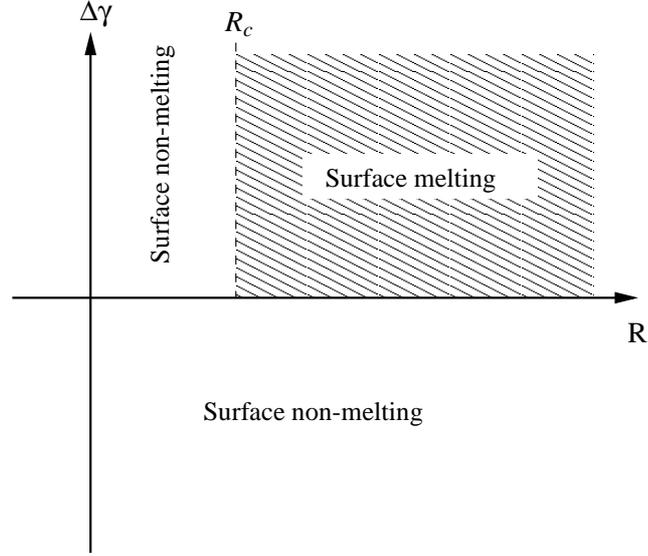}
\end{center}
\caption{\footnotesize{Schematic phase diagram for melting of small particles. The region in which surface melting occurs is shaded. For definitions of symbols, see text.}}
\label{fig:phased}
\end{figure}

 Before the solid-liquid phase transition takes place, surface melting can occur as seen in figure \ref{mprofile}b. 
When the intersection parameter $\Mp$ exceeds the crystal order parameter $M_{CRY}$ surface melting starts.  
The temperature $T_{QLL}$, for which the quasi liquid layer starts growing, is found by solving $\Mp=M_{CRY}$.Solving for  $T_{QLL}$ to first order in $R$ we get:
\beg\label{eq:qll}
\frac{T_m-T_{QLL}}{T_m}=\frac{2\Delta\gamma}{\xi L_b }+\frac{\gamsl\left(1-\frac{\Delta\gamma}{\gamsl}\right)^2}{L_b R}
\eeq
For particles with large radius $R$ the second term in equation (\ref{eq:qll}), caused by the curvature, will vanish and surface melting can only occur if $\dgam>0$. As the particle size decreases the curvature effect becomes more important. Generally, surface melting will occur when $T_{QLL}<T_c$ and when the right hand side of equation (\ref{eq:qll}) is positive.
In figure (\ref{fig:phased}) a schematic phase diagram for surface  melting is drawn. For  $\dgam>0$  there exists a critical radius $R_c$, which is given by:  \beg\label{eq:rcp}
\frac{R_c}{\xi}=\frac{1}{2}\frac{3-\kappa}{1-\kappa}\left[1+\sqrt{1-\frac{12(1-\kappa)}{3-\kappa}}\right].
\eeq
The model predicts surface melting for particles with radius $R$ larger than the critical radius $R_c$, while small particles with radius $R$ below the critical radius $R_c$ will maintain a dry surface. 
The existence of $R_c$ was also predicted from MD simulations on small Au particles by \cite{md}. They found that Au clusters containing less than 350 atoms, which corresponds to $R_c=1nm$, do not show any quasi liquid layer. \\
The critical radius $R_c$ has been measured experimentally for Pb \cite{kofmanrc} and Sn \cite{tinnano}. The experiment of \cite{tinnano} found  $R_c/\xi=5.5$ for Sn which is comparable with the model value $R_c/\xi = 3.8$.

\subsection*{Surface premelting}
For particles with $\Delta\gamma>0$ and $R>R_c$, surface melting will occur before the transition from solid to liquid. By definition of the intersection point $\Mp$, the quasi liquid layer thickness $\lqll$ is found by solving $M(r,\lqll)=\Mp$ where $M(r,\lqll)$ is given in the appendix, equation (\ref{eq:comsol}). 
The quasi liquid layer is found to grow with increasing temperature:
\beg\label{eq:lqll}
\lqll(T)=\frac{\xi}{2}\ln\left(\frac{\sigma(R)\left(1-\frac{\xi}{R}\right)}{\frac{\xi}{R}\left[1-\frac{\deltat R}{2\gamsl}\right]}\right)
\eeq
where $\sigma(R)$ is given in the appendix and $\deltat$ is given in equation \ref{eq:deltat}.
It is seen that the quasi liquid layer thickness diverges when
\beg\label{eq:qlltrans}
\deltat=\frac{2\gamsl}{R}.
\eeq

The model predicts however that the solid-liquid phase transition  takes place before the quasi liquid layer diverges, see figure \ref{mprofile}b. Equation 25 is equivalent to the Gibbs-Thomson relation which in this model gives the upper temperature limit for which a solid particle can exist. Further the quantity $2\gamsl/R$ is known as the Laplace- or capillary pressure. The pressure of a particle is always greater than the surrounding pressure by the amount $2\gamsl/R$. The quasi liquid layer therefore diverges since the difference between the particle pressure and surrounding pressure becomes to weak to keep the solid particle together.  
\section*{Discussion}
Surface pre-melting initiated from a flat surface of semi-infinite systems studied by \cite{pluis}, is found in the limit $R\rightarrow\infty$, thus equation (\ref{eq:lqll}) simplifies to :
\beg\label{eq:lqllsim}
\lqll(T)=\frac{\xi}{2}\ln\left(\frac{2\dgam}{\xi\deltat}\right),
\eeq
note that surface melting appears only for $\dgam>0$. For semi-infinite systems the quasi liquid layer thickness diverges for $T\rightarrow T_m$. This is often interpreted as the bulk melting transition.
However due to the finite size nature of any crystal the diverging quasi layer thickness is preempted by a first order melting transition. Take for example a film with thickness $L$, the melting temperature $T_c$ is given by \cite{takagi1}:
\beg
\Lambda(T_c)=\frac{1}{L_m}\frac{T_m-T_c}{T_m}=\frac{\gamsv-\gamlv}{L}.
\eeq
Thus the quasi liquid layer thickness at the melting temperature $T_c$ is of the order of $\lqll(T_c)\sim\ln(L/\xi)$.

The present work has considered free finite size particles. However due to the generality of the model the obtained results  could also be applied to coated particles or particles embedded in a matrix where both premelting \cite{kofman,johnson1} and superheating \cite{johnson2,imura} have been observed. For such systems the surface energies $\gamsv$ and $\gamlv$ should be replace by the interfacial energies, $\gamma_{sm}$(solid-coat/matrix) and $\gamma_{lm}$(liquid-coat/matrix).
\section*{Conclusion}
In this contribution, an analytic solution to the two-parabola Landau model for finite size spherical systems was given. The solution provides insight into the size effect of the melting temperature and the latent heat of fusion. For particles with size comparable to the correlation length $\xi$ a non-linear dependence on size was found for both the melting temperature and the latent heat of fusion. For large systems compared to $\xi$, classical thermodynamic results was found.  
The model further provides an analysis of surface pre-melting. It was predicted that surface pre-melting only persists on systems with radius greater than a critical size.     
\section*{Acknowledgments}
This work is supported by Danish Research Council for Nature and Universe.
J. C. would like to thank Association of International Education Japan (AIEJ), Julie Marie Vinter Hansens Fund, Nordea Danmark Fund, Julie Damms Fund and Vordingborg Gymnasiums fund for financial support during this work. Furthermore J.C. is grateful to professor H. Saka at Nagoya University for the hospitality during this work.
\section*{Appendix}
The general order parameter profile $M(r)$ including surface pre-melting is given in this appendix. For simplicity $\xi$ is taken as unit length. $\lqll$ signify the quasi liquid layer thickness and $\ell$ is defined by: $\ell=R-\lqll$.
\begin{equation}\label{eq:comsol}
M(r,\lqll)=
        \begin{cases}
 M_{CRY}(r)=1-\frac{1}{1+\epsilon}\frac{\ell}{r}\frac{\sinh(r)}{\sinh(\ell)} & \text{for $M>\Mp$}\\[3mm]
 M_{QLL}(r)=\frac{\beta\ell}{r}\exp(r-\ell)\left\{\sigma+\exp(2[R-r])\right\} & \text{for $M<\Mp$}
        \end{cases}
\end{equation}
where
\beg
\epsilon=\frac{\left[\coth(\ell)-\frac{1}{\ell}\right]\left[\sigma+\exp(2\lqll)\right]}{\exp(2\lqll)\left(1+\frac{1}{\ell}\right)-\sigma\left(1-\frac{1}{\ell}\right)}
\eeq
\beg
\sigma=\frac{\kappa\left(1+\frac{\xi}{R}\right)-1}{\kappa\left(1-\frac{\xi}{R}\right)+1}
\eeq
and
\beg
\beta=\frac{\epsilon}{1+\epsilon}\frac{1}{\sigma+\exp(2\lqll)}.
\eeq

\clearpage

\end{document}